# Pneumonia Detection in Chest X-Rays using Neural Networks

Narayana Darapaneni[1], Ashish Ranjan[2], Dany Bright[3], Devendra Trivedi[4], Ketul Kumar[5], Vivek Kumar[6], and Anwesh Reddy Paduri[7]

*[1]Northwestern University/Great Learning, US*

*[2-7] Great Learning, Bangalore, India*

*anwesh@greatlearning.in*

***Abstract:*** With the advancement in AI, deep learning techniques are widely used to design robust classification models in several areas such as medical diagnosis tasks in which it achieves good performance. In this paper, we have proposed the CNN model (Convolutional Neural Network) for the classification of Chest X-ray images for Radiological Society of North America Pneumonia (RSNA) datasets. The study also tries to achieve the same RSNA benchmark results using the limited computational resources by trying out various approaches to the methodologies that have been implemented in recent years. The proposed method is based on a non-complex CNN and the use of transfer learning algorithms like Xception, InceptionV3/V4, EfficientNetB7. Along with this, the study also tries to achieve the same RSNA benchmark results using the limited computational resources by trying out various approaches to the methodologies that have been implemented in recent years. The RSNA benchmark MAP score is 0.25, but using the Mask RCNN model on a stratified sample of 3017 along with image augmentation gave a MAP score of 0.15. Meanwhile, the YoloV3 without any hyperparameter tuning gave the MAP score of 0.32 but still, the loss keeps decreasing. Running the model for a greater number of iterations can give better results.

## 1. Introduction

Children under the age of five are more likely to die of pneumonia than any other cause. Pneumonia is responsible for more than 15% of all fatalities in children under the age of five worldwide. Because it can be caused by viruses, bacteria, or fungus, the researchers focused their research on applying deep learning algorithms to diagnose pneumonia based on chest X-ray images. Pneumonia caused over 500,000 visits to emergency rooms and over 50,000 deaths in the United States in 2015, maintaining it in the top ten causes of mortality in the country. Simple actions can be used to prevent illness, and low-cost, low-tech medicine and care may be used to treat it.

Pneumonia diagnosis is a time-consuming process that involves highly skilled professionals to analyse a chest radiograph or chest X-ray (CXR) and confirm the diagnosis with clinical history, vital signs, and laboratory tests. It helps doctors to work out the extent and placement of the infection in the lungs. Respiratory illness manifests as a neighbourhood of inflated opacity on X-Ray. However, the identification of respiratory illness in CXR is troublesome due to different conditions which may appear as opacity in the lungs such as - carcinoma, bleeding, pulmonary edema, etc. The CXRs of the patient that are taken at different intervals and the correlation with clinical symptoms are useful in identifying pneumonia.

A chest X-ray examines your lungs, bones, and heart using a focussed beam of radiation. Chest X-rays are quick, non-invasive tests. Usually, one will know the results of their X-ray within one to two days. Faster diagnosis can guarantee timely access to treatment and save much needed time and money. Chest X-rays use focused beams of radiation. The images created by these radiation beams are of the inside of your body. The negative pictures of black-and-white photographs resemble X-ray images. The thickness of your body's tissues varies. Every part of your body has a different ratio of radiation passing through it. Your bones, for instance, are incredibly thick and don't let much radiation through. On an X-ray picture, bones appear white. Your lungs, on the other hand, enable more radiation to get through. An X-ray picture of your lungs shows them to be grey. To identify and treat health disorders, healthcare practitioners examine the colours and shading on an X-ray.
Chest X-rays are taken by a radiology technician. X-ray testing is a specialty for these professionals. The appearance of the CXR is affected by factors such as the patient's posture and the depth of inspiration, making it more difficult to interpret. In addition, radiologists must interpret a large number of CXRs on a daily basis. This repetitive task and high-end expertise of clinicians can be minimized with the potential of AI and ML to automate the initial detection or image screening of potential respiratory illness cases to grade and expedite their review.
Building an algorithm to automatically identify whether a patient is suffering from pneumonia or not bypassing the x-ray to the model should be extremely accurate because people's lives are at stake. In this study, we tried to build a model that classifies and localizes the pneumonia infection in the lungs. We have tried various possible algorithms on limited computational resources by downsampling, downsizing the images, etc. Our study aims to achieve the benchmark results by combining various methods that achieve greater accuracy with less downsampled and downsized images. This method will help us to experiment and deliver better results with less computational resources and less data which will eventually help us to build a better-advanced computer vision algorithm that destroys the barrier of resources needed for computation.

### *1.1. Literature Review*

Pneumonia is a lung illness caused by an acute respiratory infection. 1 out of 5 deaths in children occurs due to pneumonia [1]. The number of visits to emergency departments with pneumonia as the primary diagnosis is 1.5 million in the US (Stats as per CDC in 2018). In 2018, India reported the second-highest incidence of pneumonia-related fatalities among children under the age of five. The majority of deaths occurred in children under the age of two, with about 153,000 deaths occurring during the first month of life. It can be prevented with simple interventions and treated with low-cost, low-tech medication and care. Pneumonia is usually confirmed once taking a Chest X-ray. It helps doctors to work out the extent and placement of the infection. respiratory illness sometimes manifests as a neighbourhood of inflated opacity [2] on X-Ray. However, the identification of respiratory illness in CXR is troublesome due to different conditions which may appear as opacity in the lungs such as - carcinoma, bleeding, pulmonary oedema etc. The CXRs of the patient that are taken at different intervals and the correlation with clinical symptoms are useful in identifying pneumonia. Factors like the positioning of the patient and depth of inspiration change the appearance of the CXR [3], making the interpretation more difficult. Moreover, radiologists have to interpret the high volume of CXRs every day. This repetitive task and high-end expertise of clinicians can be minimized with the potential of AI and ML to automate the initial detection or image screening of potential respiratory illness cases to expedite their review. In the medical domain, most people use CNN's transfer learning method for classification with an adequate dataset. Okeke S. et al. [10] performed a study on the large dataset containing the train, validation & test, and two sub-folders include Pneumonia and non-Pneumonia class. The proposed model was made from scratch to the end that extracts into two parts: a feature extractor & a classifier. In the pre-processing part, data augmentation was applied. The model consisted of several Conv. layers with max-pooling and Relu activation. During the training period, different types of output sizes were tested and combined into a 1D feature for the classifier procedure. While the classifier, the dropout (0.5), 2 dense layers with Relu and output with the Sigmoid activation were implemented. The final results were that, among several image shapes, the best shape was (200 x 200 x 3) and it has a validation score - 94% with a loss of 0.18%. The study encountered that the larger the image size, the less validation score, vice versa, the smaller size images have performed well. Pan et al. performed a two-step classification [4]. For classification, an ensemble of 3 transfer learning models was used which were trained on ImageNet. For object detection, an ensemble of 2 models was used. The image was first classified as positive or negative. In the second stage, the object detectors were used to form the bounding boxes. The object detection model was trained only on positive images. The Metadata was also integrated however, any relevant inference for the model building was not found. The original image resolution of the RSNA dataset (1024x1024) was reduced and several resolutions (256, 320, 384, 448, 512) were tried. Training on small size image resolution did not result in any decrease in performance. The data were also stratified into ten folds. The predicted bounding box was reduced by 12.5% in size to achieve better accuracy on the test data. Cheng et al. implemented a single stage architecture that performed both object classification and bounding box localization [5]. The X-rays were reduced to a lower resolution of 224x224. The training dataset was increased using image augmentations like - random rotations, translations, scaling, horizontal flipping, and addition or subtraction of random constants. The dataset was divided into a 95% training set and a 5% validation set. RetinNet pre-trained on the ImageNet dataset was used to classify the images. Arman H. et al. presented a paper [11] using a COVID-CXNet while collecting several datasets and combining them into one. In the Pre-processing, the images were augmented, normalized, resized into (320 x 320) and for clarity- CLAHE & BEASF were being used. Three types of experiment performed, In the base model- simple architecture contains simple architectures with no BN and Pooling. Second exp.- ChexNet built with DenseNet has been performed and the third exp.- the COVIDNet implemented where the CheXNet model used as a backbone. The Grad-Cam technique was used for the evaluation. And the results were explained where the base model got an AUC of 0.99, 98% accuracy and 0.94 F1 scores achieved. The second experiment over-fitted and got bad results and the third experiment- got 87.2% accuracy. In the end, the study claimed that the proposed model enhances the performance and reduces the overfitting and for the F1 score; it has a few positive specimens that tend to overfit. Joseph P. et al. [12] delivered a model of COVID-19 pneumonia for a frontal chest x-ray tool for monitoring the disease without invasive. The study used two types of dataset- i) COVID-19 Cohort ii) Non-COVID-19 dataset (blended multiple data like RSNA etc., into one). For the cohort dataset, unknown radiologists extracted the information with the following scores- (1) ground-glass opacity/consolidation for particular lungs as 0 = no involvement, and 1 = < 25%, 2 = < 50%, 3 = < 75%, 4 = > 75% were considered as involvement with the total score in ranged from 0 to 8 (2) degree of opacity for particular lungs where 0 = no opacity, 1 = ground glass, 2 = consolidation, 3 = white-out with the total score in the range of 0 to 6. The image was shaped into (224 x 224) and scaled to [-1024, 1024]. A DenseNet model was implemented because the study found that it worked well to predict Pneumonia disease. The experiments were performed on the four categories- intermediate feature, 18 outputs, 4 outputs and lung opacity outputs. Among them, the best model which has fewer features (lung opacity) achieved a good score followed by MAE–0.78 and MSE 0.86. In the end, the team stated that fewer features and parameters work best to prevent overfitting. So, the proposed model may be used for de-escalation/escalation especially in ICU. Alberto S. et al. [13] made an end-to-end model on CXR images for lung detection of COVID-19 patients. The trained model used weakly supervised learning methods like– segmentation, score estimation, and spatial alignment on the different types of datasets. In that case, the study demonstrated the BS-Net model, which gives high accuracy of all stages of pre-processing. The team has collected different kinds of datasets– Montgomery Country, Shenzhen Hospital, JSRT database, Brixia COVID-19 dataset, and public COVID-19 dataset. During the implementation, different architectures like-ResNet, VGG, DenseNet and Inception were used as backbone models. In the segmentation part, the U-Net++ architecture extracts the lung's shape and edges in the alignment part. The Brixia prediction scores were used with the help of the Ensemble Decision Strategy (ENS). The achieved scores were as follows: MAE of 0.041 and 0.0424. The study informed that the score of +/- 0.5 for the clinical perspective and +/- 2 global perspective are acceptance scores. In the end, the model obtained the details of robustness and passed the performance in consistency and accuracy. Deep learning's image processing technique impacts positivity on the diagnosis of X-ray vision. Rachna J. et al. [14] have collected the dataset from Kaggle containing images related to COVID-19 positive, normal and pneumonia. The dataset has been split into training–5467, validation–965. The input shaping (128 x 128 x 3) has been used. The project was performed on different types of architecture like– Inception V3, Xception and ResNet. In the pre-processing part, image augmentation to avoid over-fitting. During the performance, it was found that if the image shape- 128, then it will train faster, and if the image shape- 256, then it will perform well. Hence, 256 input shapes for XCeption, Inception and 128 for ResNeXt, and applied LeakyRelu as activation function. A classification report was used for the evaluation. Among the several models, Xception Net achieved an excellent score of– 0.97. In the end, the study claimed the model performed well in the testing but maybe it results from over-fitting, thus it can verify against the new public dataset where it can achieve the true accuracy. Earlier, the patients found abnormalities in the chest X-ray that denotes infection of COVID-19. Utilizing the above idea, Linda W. et al. [15] presented a COVID-Net model to detect X-ray images. Creating a dataset, several datasets like RSNA Pneumonia merged, and the result of a dataset called COVIDx dataset. This dataset has 13,975 CXR images, and it is now open for public access and it is classified into three categories such as 1) no infection, 2) non-COVID-19 infection 3) COVID-19 viral infection. For the pre-processing part, they cropped 8% of the image and in the image augmentation– they initialized translation (10%), rotation (10%), flip, zoom (15%) and shift (10%) etc. The COVID-Net is trained on the ImageNet dataset, by following architecture VGG and DenseNet. Also, the study implemented Adam as an optimizer, and following hyper-parameters as

learning = 2e-4, epochs = 22, factor = 0.7 and patience = 5 and heterogeneous convolutional layers and various kernels range from 1 to 7. The results show that COVID-Net architecture outperforms the other models. The accuracy score is 93.3%, sensitivity score is 91%. In the end, the team has determined that this architecture is more lucid and dependable during the screening process and still gives rapid evaluation. Dingding W. et al. [16] came up with an idea to reduce the burden of X-ray imaging diagnoses. First, for an experimental purpose, five transfer learning methods [Xception, VGG16, Inception V3, ResNet50 and DenseNet121] were used to obtain the information. Then the team proposed a method to improve the screening diagnostic and accuracy. The method is to blend the neural network and machine learning to make an end-to-end model. In the study, the combined datasets were combined and used a re-sampling method to overcome the unbalance problem. In the pre-processing part, the input shape is taken as (224 x 224), plus augmentation and normalizing the image between [-1, 1] to avoid over-fitting. Custom transfer learning is used as the base model at first. Then, the second method is applied to the bottleneck feature, where the pre-trained model is taken as the feature extraction and as a result, extracted features are a lower dimension vector to reduce time. During the training, hyper-parameters–learning rate = 1e-7, epochs = 1000, batch size = 32 have been implemented. In the end, the saved bottleneck feature taken as the input of the machine learning algorithms [DT, RF, AdaBoost, Bagging, SVM] predicted the COVID-19 images. Sensitivity, Specificity and Accuracy have been used for the evaluation. The results of the base model- Xception architecture- 94.16% sensitivity, specificity- 99.17%, the F1- 96.38% and the AUC- 96.54%. Among the several combined bottleneck methods, the best model was Xception + SVM. The results of this model were sensitivity- 99.27%, specificity– 99.38%, F1– 96.63% and AUC– 96.77%. Therefore, the researcher concludes that the blended feature of deep learning and machine learning gives more efficiency than the conventional methods. Contemporary diagnosis methods are common visualisation techniques for x-ray likewise antigen test/molecular for identifying the severity and presence of the disease. Certain ranges of age play a crucial role in classifying the patients. However, several models trained on the RSNA dataset, which particularly contains the images of the children class. Thus these models may give biased results on certain cases of the disease. That's why Julian D. et al. [17] proposed the case study to show how much pre-processing work gives accurate results and compares it with the other models. In the research, two sets of images - Posterior Anterior & Anterior-Posterior in the XR format have been aggregated together. The dataset is divided into three parts- control, Pneumonia, COVID-19. In the image pre-processing part, XR images were converted into uncompress 16 bits PNG format and then applied to the Monochrome 2 filter without losing the details. The proposed model is COVID-Net. The image downsized to (224 x 224 x 3), normalized with the mean of [0.485, 0.456, 0.456], including std of [0.229, 0.224, 0.225] with the Z-Score function. The hyper-parameters were- lr= 2e-5, epochs = 24, batch_size = 32, factor = 0.5 and patience = 3 implemented. Image augmentation- flip, Gaussian noise plus variance- 0.015 and scaling applied. Afterwards, Relu and dropout (0.3) were implemented in the first stage of FC layers. The three types of methodology were applied. In the first method, only raw data without implementing the filtering and augmentation and with the original size used. In the second method, the data were pre-processed by zooming, cropping, and resizing into a square image of (1:1 ratio). The main part is masking where lungs are separated from the U-Net segmentation. And in the last, greyscale was added to the vertical & horizontal levels. That identifies the squared area of the lungs and creates a histogram from it. Finally, the last method was almost the same as the second method. The only differences were that an extra kernel (5 x 5) was added to make it superimposed and wider only to the chest region. As a result, method 1 performs better with AUC- 0.99 and global performance accuracy- 91%. Method 2- performed poorly than method 1 with ROC- 81% - 93%, and the global performance accuracy- 87%. And last, method 3, is slightly worse than method 2 in terms of COVID-19 class. While it got a global score of 0.98%. Several empirical studies have worked on the expert Dataset by the Stanford ML Group which contains 224,316 X-rays scans of 65,240 patients, where each report was labelled for the presence of 14 observations as positive, negative, or uncertain.Common findings from the works done by H.H.Pham et al. (2019) and Z. Yuan et al. (2020) were, for better handling of the uncertain observations which occupy a significant portion of many CXR datasets H.H.Pham et al. (2019) proposed to use label smoothing technique and is used in the work of Z. Yuan et al. (2020). In the inference stage, the test-time augmentation (TTA) and data augmentation were applied for each CXR respectively. The model was trained on a different transfer learning algorithm and it was observed that different classifiers yield different scores for each label. Hence an ensemble technique is explored to achieve a highly accurate classifier. Following six architectures were used namely DenseNet-121, DenseNet-169, DenseNet-201, Inception-ResNet-v2, Xception, and NASNetLarge. The output of all the above-trained networks was taken to make the ensemble model which gave an AUC score of 0.930 for [6] H.H.Pham et al. (2019). To improve the AUC score, the AUC margin-based surrogate loss function was proposed by Z. Yuan et al. (2020) over the AUC square loss which is generally used. A Two-stage Training Framework for DAM was used where for the pre-training stage, the focus is on learning by optimizing the standard cross-entropy loss on different transfer learning algorithms like DenseNet121, DenseNet161, DensNet169, DensNet201 and Inception-ResNet-v2. In the second stage, the focus is on AUC maximization by fine-tuning the decision boundary of the classifier by optimizing the AUC margin loss. The last classifier layer trained weights in the first stage are replaced by random weights and the DAM method is used to optimise the last classifier layer and all previous layers. Subsequently, an AUC score of 0.9305 was achieved which puts Z. Yuan et al. (2020) [7] work at the top of the CheXpert competition. Mohammad Farukh Hashmi et al[18] experimented on a Large Dataset of Labeled Optical Coherence Tomography (OCT) and Chest X-Ray Images[8] which has 5858 images. Dataset was labelled by two physicians and evaluated by another physician. 5156 images used the training set and 700 images in the test set. Viral and bacterial pneumonia is considered pneumonia infected. Normal/no pneumonia are augmented twice to remove imbalance data in the training set. Different neural networks expect images of different sizes according to their defined architecture, so images were resized to 224*224 & 229*229. All 5 models i.e. ResNet18, DenseNet121, and MobileNetV2, InceptionV3 and Xception were trained individually and then a weighted sum was taken to classify. SGD as the optimizer was used as it has better generalization[9], and the model was trained for 25 epochs. The learning rate, the momentum, and the weight decay were set to 0.001, 0.9, and 0.0001, respectively. Case1 - Every model contributed equally towards the final predictions i.e 0.20 Weight of each architecture was considered. Test accuracy of 97.45 and a loss of 0.087 was obtained. In case2-optimum weights were taken based on individual architecture classification performance, found out the final weighted classifier was able to achieve a testing accuracy of 98.43, and the testing loss was 0.062 with an AUC of 99.76. M. E. H. Chowdhury et al. [19] conducted experiments on deep networks such as ResNet101, VGG19, DenseNet201, CheXNet, and InceptionV3 and shallow networks such as ResNet18, SqueezeNet, MobileNetv2.CXR's were normalised and resized as their requirement for pre-defined architecture like for SqueezeNet, the images were resized to 227 × 227 pixels, for Inceptionv3 the CXR's were resized to 299×299 pixels and for mobilenetV2, ResNet18, ResNet101, VGG19 and DenseNet201, the CXR's were resized to 224 × 224 pixels. In Case1, COVID-19 chest X-ray images were significantly less than 423, images were downsampled to balance features, so equal samples were randomly selected from normal (out of 1579) and viral pneumonia (out of 1485) to balance data. In case2, all CXR (i.e., 423 COVID-19, 1579 normal and 1485 viral pneumonia images) were used. Due to an imbalance in the number of features COVID-19 images were increased sixfold while normal and viral pneumonia images one time to maintain the balance between classes by augmenting. Each study was evaluated using a stratified 5-fold cross-validation with a ratio of 80:20 for training: testing, where 10% of training data was used as a validation set to test hyperparameter and avoid overfitting.

## 2. Introduction to Data and Pre-Processing

The dataset is collected from the following website: RSNA Pneumonia Detection Challenge | Kaggle. With almost 54,000 members from 146 countries, the Radiological Association of North America (RSNA) is an international society of radiologists, medical professionals, and other medical physicists.They propose that machine learning can help prioritise and accelerate the assessment of possible pneumonia cases by optimizing initial detection (imaging screening). RSNA shared 30000 plus x-ray images and its diagnoses information along with the bounding box coordinates in two different .csv files for the Kaggle competition. Every x-ray image is in Digital Imaging and Communications in Medicine (DICOM) format which is the standard medical imaging format accepted worldwide. It is a format that has metadata, as well as pixel data or image data attached to it. Therefore, each image has metadata information like Patient ID, name, age and other image-related data.

### *2.1. Sample Data*

RSNA - CXR Dataset contains 30227 X-ray images in Dicom format. There are three classes with 31.61% lung opacity, 39.11% -no lung opacity, 29.28% normal images. Thus, in the target class, there are 31.61% of pneumonia class, 68.38% of non-pneumonia images. There are 3543 duplicate entries. Some of them are different X-ray views for the same patient. Bounding boxes for patients having pneumonia are defined in the train labels file. There are 9555 positive patients in this file. Each X-ray has metadata associated with it. It gives information about the patient, the view position etc. There are 3543 duplicate entries.

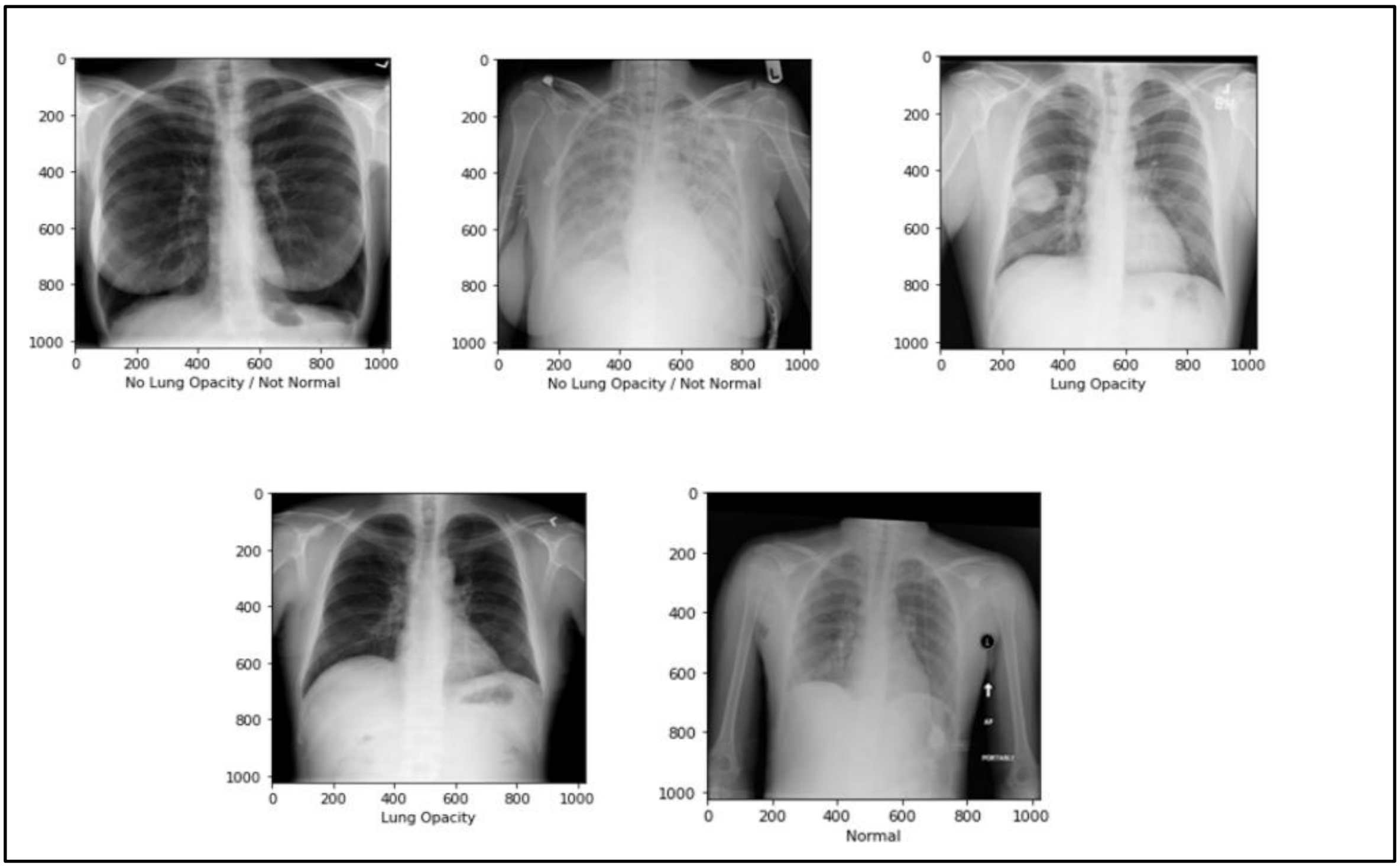

Fig. 1. Sample Images of chest X-rays

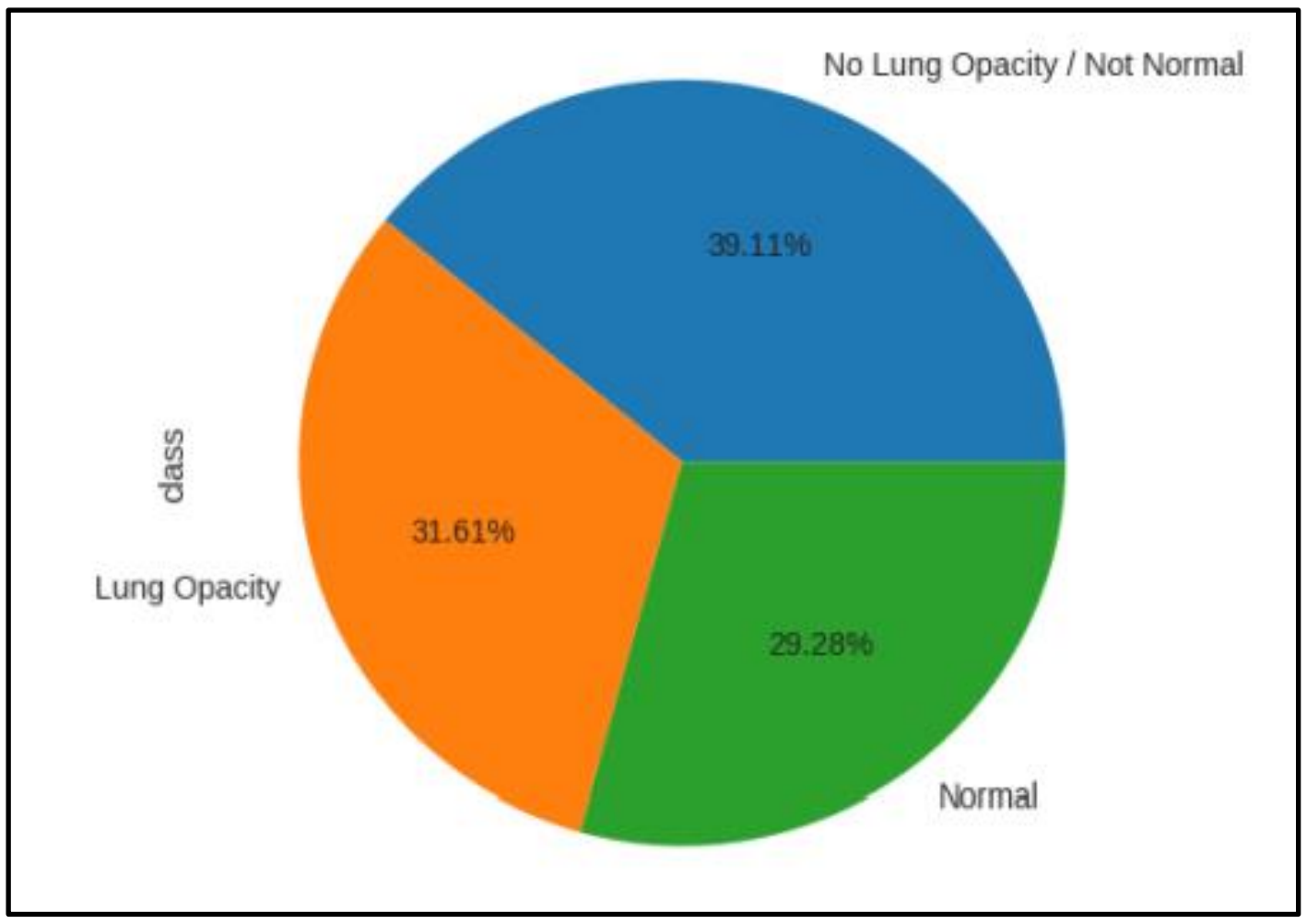

Fig. 2. Pie chart of three classes

Stratified sampling ensures the diversity of the sample and the overall variance in the population. For the RSNA dataset, the minimum stratified sample according to Cochran's formula would be the following.

| | class | size | samp_size |
|---|---|---|---|
| 0 | Lung Opacity | 9555 | 703 |
| 1 | No Lung Opacity / Not Normal | 11821 | 870 |
| 2 | Normal | 8851 | 651 |

Fig. 3. Stratified samples

## *2.2. Exploratory Data Analysis*

### *2.2.1. Diagnose CSV file*

| | patientId | class |
|---|---|---|
| 0 | 0004cfab-14fd-4e49-80ba-63a80b6bddd6 | No Lung Opacity / Not Normal |
| 1 | 00313ee0-9eaa-42f4-b0ab-c148ed3241cd | No Lung Opacity / Not Normal |
| 2 | 00322d4d-1c29-4943-afc9-b6754be640eb | No Lung Opacity / Not Normal |
| 3 | 003d8fa0-6bf1-40ed-b54c-ac657f8495c5 | Normal |
| 4 | 00436515-870c-4b36-a041-de91049b9ab4 | Lung Opacity |

Fig. 4. Sample of dataset

### *2.2.2. Bounding Box Information*

In the bounding boxes dataset, there are Null values for x, y, width and height columns. As these patients are not having pneumonia there is no bounding box information required for such patients. Therefore, those values are Null. During object detection, the Null values will be omitted.

| | patientId | x | y | width | height | Target |
|---|---|---|---|---|---|---|
| 0 | 0004cfab-14fd-4e49-80ba-63a80b6bddd6 | NaN | NaN | NaN | NaN | 0 |
| 1 | 00313ee0-9eaa-42f4-b0ab-c148ed3241cd | NaN | NaN | NaN | NaN | 0 |
| 2 | 00322d4d-1c29-4943-afc9-b6754be640eb | NaN | NaN | NaN | NaN | 0 |
| 3 | 003d8fa0-6bf1-40ed-b54c-ac657f8495c5 | NaN | NaN | NaN | NaN | 0 |
| 4 | 00436515-870c-4b36-a041-de91049b9ab4 | 264.0 | 152.0 | 213.0 | 379.0 | 1 |

Fig. 5. Sample of a bounding box

### *2.2.3. Dicom Metadata*

The inferred file was captured in the DCM format. Few important details from the metadata are: the body part examined - ‘Chest’, View Position - ‘AP/PA’, Patient’s Sex - ‘M/F’, Photometric Interpretation - ‘Monochrome’,

Rows - 1024, Columns - 1024, Bits Allocated - 8, Conversion Type - ‘ WSD’ (Workstation), Modality - ‘CR’ (Computed Radiography).

```
Dataset.file_meta -------------------------------
(0002, 0000) File Meta Information Group Length  UL: 200
(0002, 0001) File Meta Information Version       OB: b'\x00\x01'
(0002, 0002) Media Storage SOP Class UID         UI: Secondary Capture Image Storage
(0002, 0003) Media Storage SOP Instance UID      UI: 1.2.276.0.7230010.3.1.4.8323329.6379.1517874325.469569
(0002, 0010) Transfer Syntax UID                 UI: JPEG Baseline (Process 1)
(0002, 0012) Implementation Class UID            UI: 1.2.276.0.7230010.3.0.3.6.0
(0002, 0013) Implementation Version Name         SH: 'OFFIS_DCMTK_360'
-------------------------------------------------
(0008, 0005) Specific Character Set              CS: 'ISO_IR 100'
(0008, 0016) SOP Class UID                       UI: Secondary Capture Image Storage
(0008, 0018) SOP Instance UID                    UI: 1.2.276.0.7230010.3.1.4.8323329.6379.1517874325.469569
(0008, 0020) Study Date                          DA: '19010101'
(0008, 0030) Study Time                          TM: '000000.00'
(0008, 0050) Accession Number                    SH: ''
(0008, 0060) Modality                            CS: 'CR'
(0008, 0064) Conversion Type                     CS: 'WSD'
(0008, 0090) Referring Physician's Name          PN: ''
(0008, 103e) Series Description                  LO: 'view: AP'
(0010, 0010) Patient's Name                      PN: '00436515-870c-4b36-a041-de91049b9ab4'
(0010, 0020) Patient ID                          LO: '00436515-870c-4b36-a041-de91049b9ab4'
(0010, 0030) Patient's Birth Date                DA: ''
(0010, 0040) Patient's Sex                       CS: 'F'
(0010, 1010) Patient's Age                       AS: '32'
(0018, 0015) Body Part Examined                  CS: 'CHEST'
(0018, 5101) View Position                       CS: 'AP'
(0020, 000d) Study Instance UID                  UI: 1.2.276.0.7230010.3.1.2.8323329.6379.1517874325.469568
(0020, 000e) Series Instance UID                 UI: 1.2.276.0.7230010.3.1.3.8323329.6379.1517874325.469567
(0020, 0010) Study ID                            SH: ''
(0020, 0011) Series Number                       IS: "1"
(0020, 0013) Instance Number                     IS: "1"
(0020, 0020) Patient Orientation                 CS: ''
(0028, 0002) Samples per Pixel                   US: 1
(0028, 0004) Photometric Interpretation          CS: 'MONOCHROME2'
(0028, 0010) Rows                                US: 1024
(0028, 0011) Columns                             US: 1024
(0028, 0030) Pixel Spacing                       DS: [0.139, 0.139]
(0028, 0100) Bits Allocated                      US: 8
(0028, 0101) Bits Stored                         US: 8
(0028, 0102) High Bit                            US: 7
(0028, 0103) Pixel Representation                US: 0
(0028, 2110) Lossy Image Compression             CS: '01'
(0028, 2114) Lossy Image Compression Method      CS: 'ISO_10918_1'
(7fe0, 0010) Pixel Data                          OB: Array of 119382 elements
```

Fig. 6. Metadata information

### *2.2.4. Summary*

1. The target column has 2 unique values and its distribution is as follows: -

```
0    20672
1     9555
Name: Target, dtype: int64
```

Fig. 7. Distribution of normal and positive classes

2. 1/3rd of the patients have pneumonia.
3. Pneumonia is confirmed by the presence of opacity in the chest X-ray which is why No Lung Opacity / Not normal class is considered not pneumonia.
4. We have 3543 duplicated Patient IDs. Upon further analysis, it is found that some x-rays have multiple bounding boxes due to which Patient IDs are repeated to save the multiple bounding box information.

### 2.2.5. Correlations

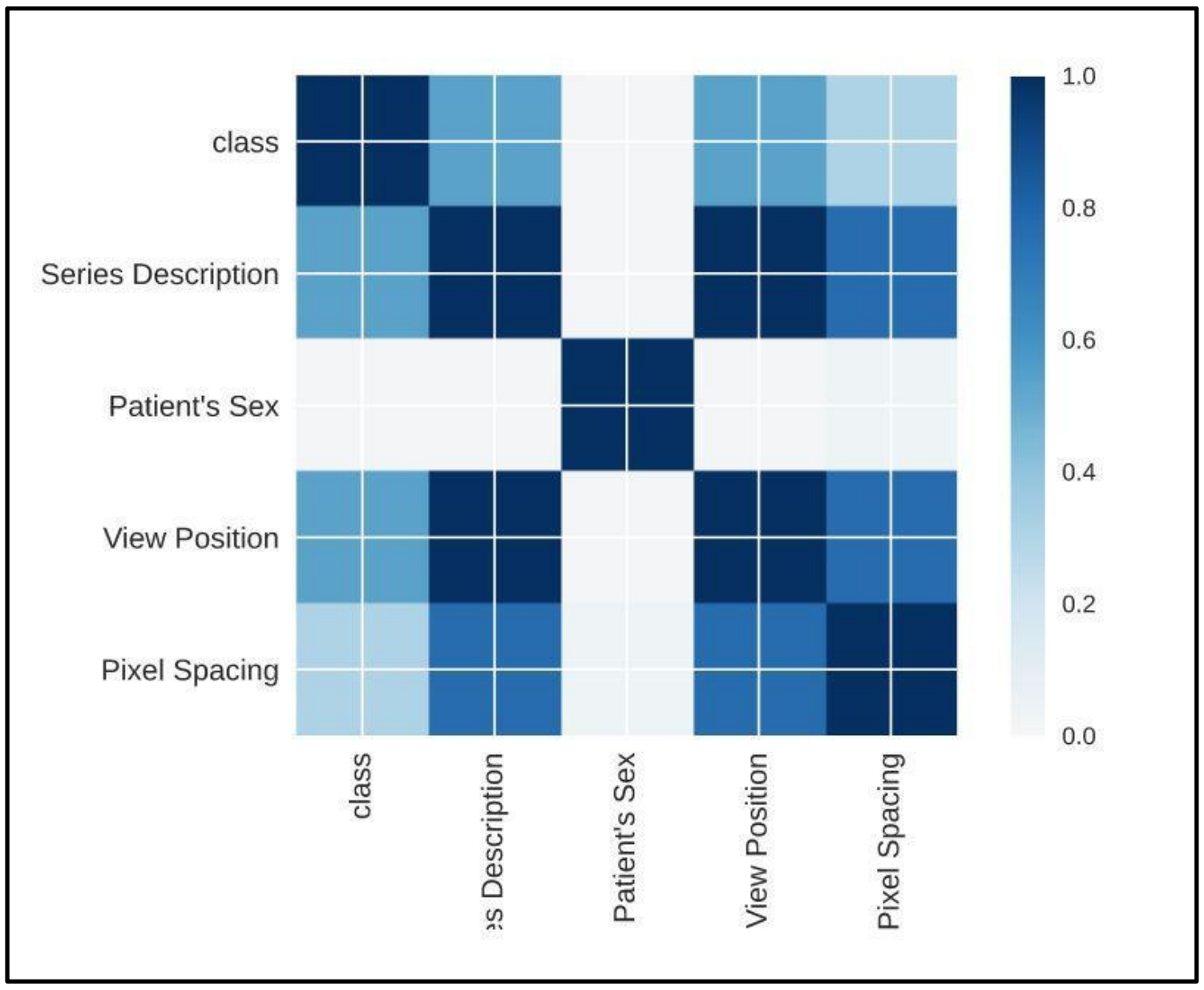

Fig. 8. Correlation with the feature

1. Only Series Description, Patient's Sex, View Position and Pixel Spacing has some relation with the class attribute.
2. Series Description and View Position are having the same values but are represented differently hence either of the two attributes can be used towards modelling.
3. From the total of 40 attributes, only 4 attributes are having some contributing factors towards classification.
4. We can use age, sex, view position and pixel spacing as inputs to form the penultimate layer of the classifier to improve performance.

### 2.2.6. Visualization

#### 2.2.6.1. Univariate Analysis

##### 2.2.6.1.1. Patient's Sex

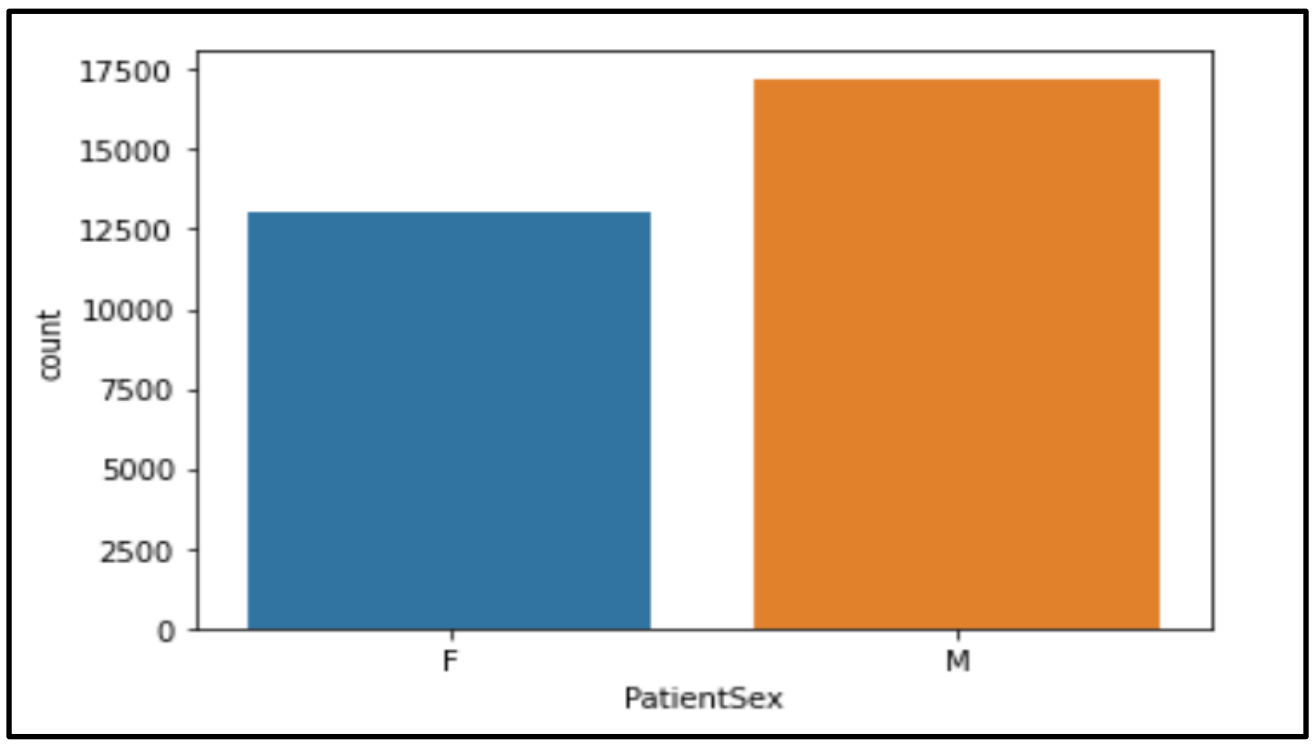

Fig. 9. The ratio of male and female

Inference: More male patient chest X-rays are present compared to the female counterpart.

*2.2.6.1.2. Patient's Age*

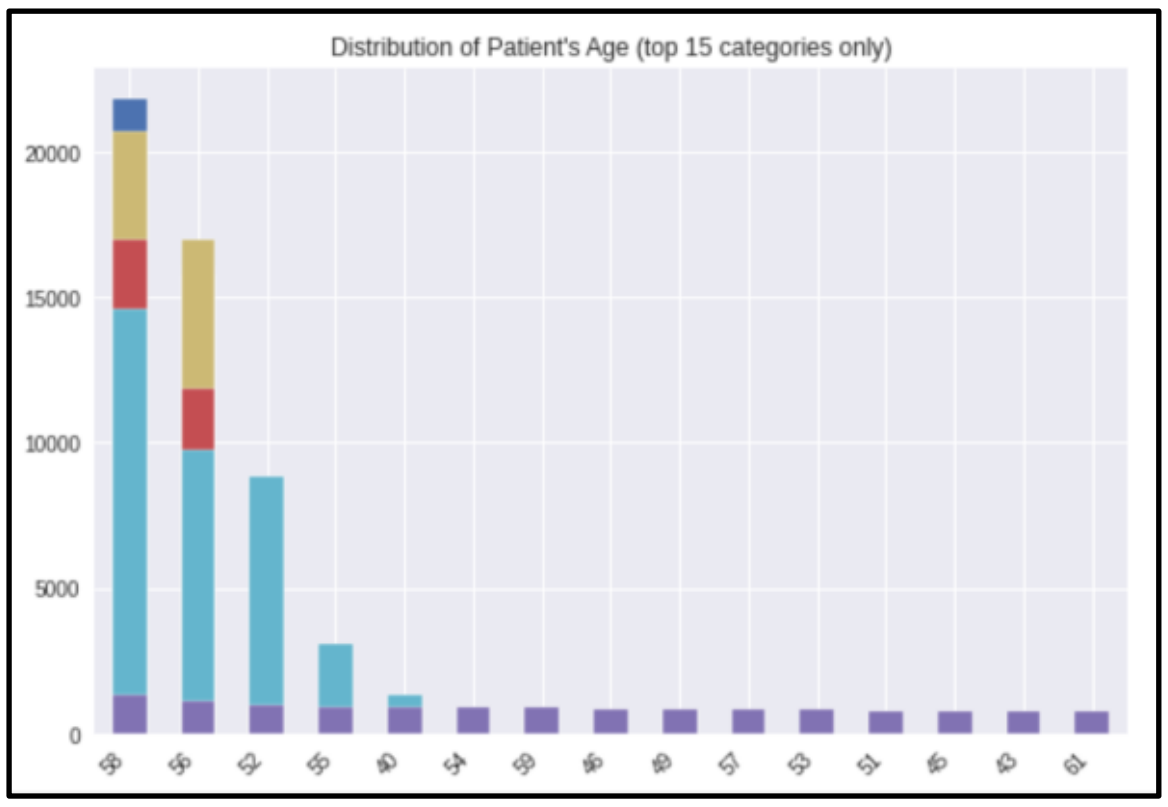

Fig. 9. Bar chart of age

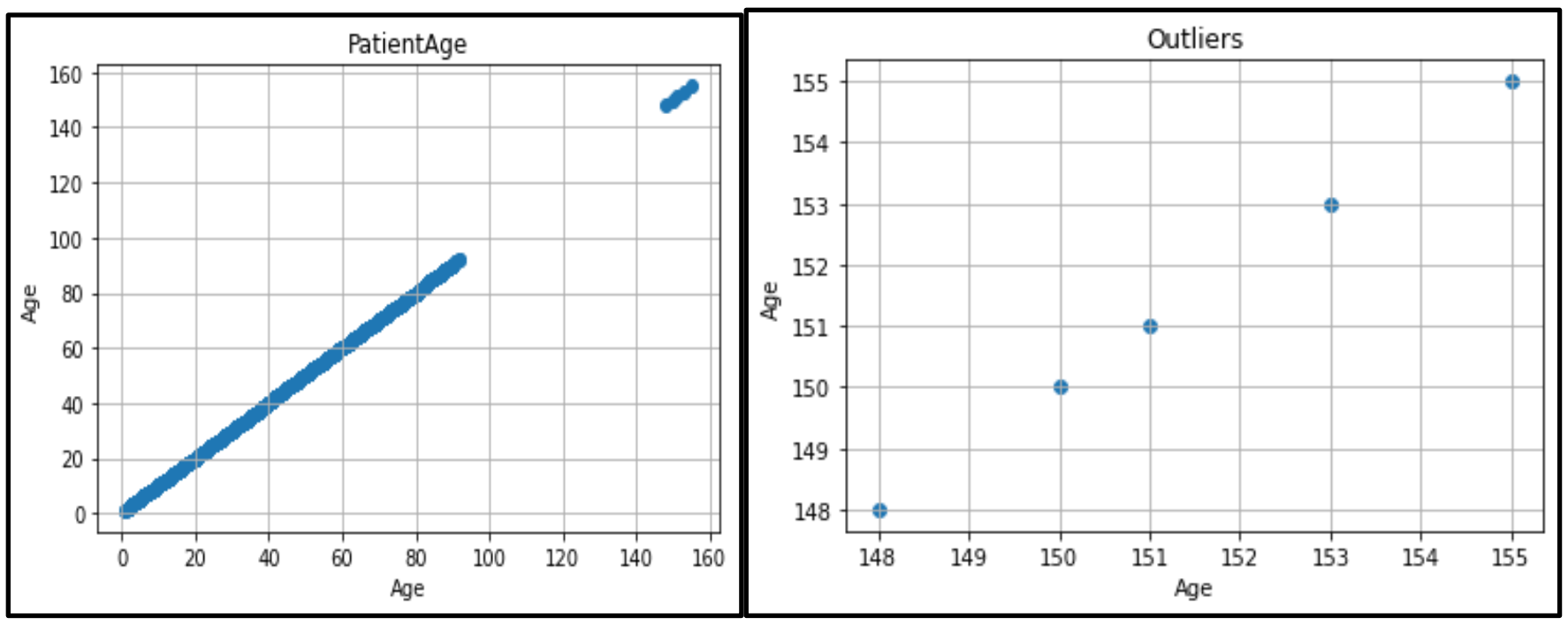

Fig. 10. (a). Scatter plot of age; (b). Detected outliers

The age distribution is between 0-100 years except for the outliers which can be due to data entry mistakes. Only 5 outliers' values are present which is above 148 years. Most patients are in the range of 55-58 years whereas maximum patients are 58 years old. The other concentrated range is between 40-54 years.

*2.2.6.1.3. View Position*

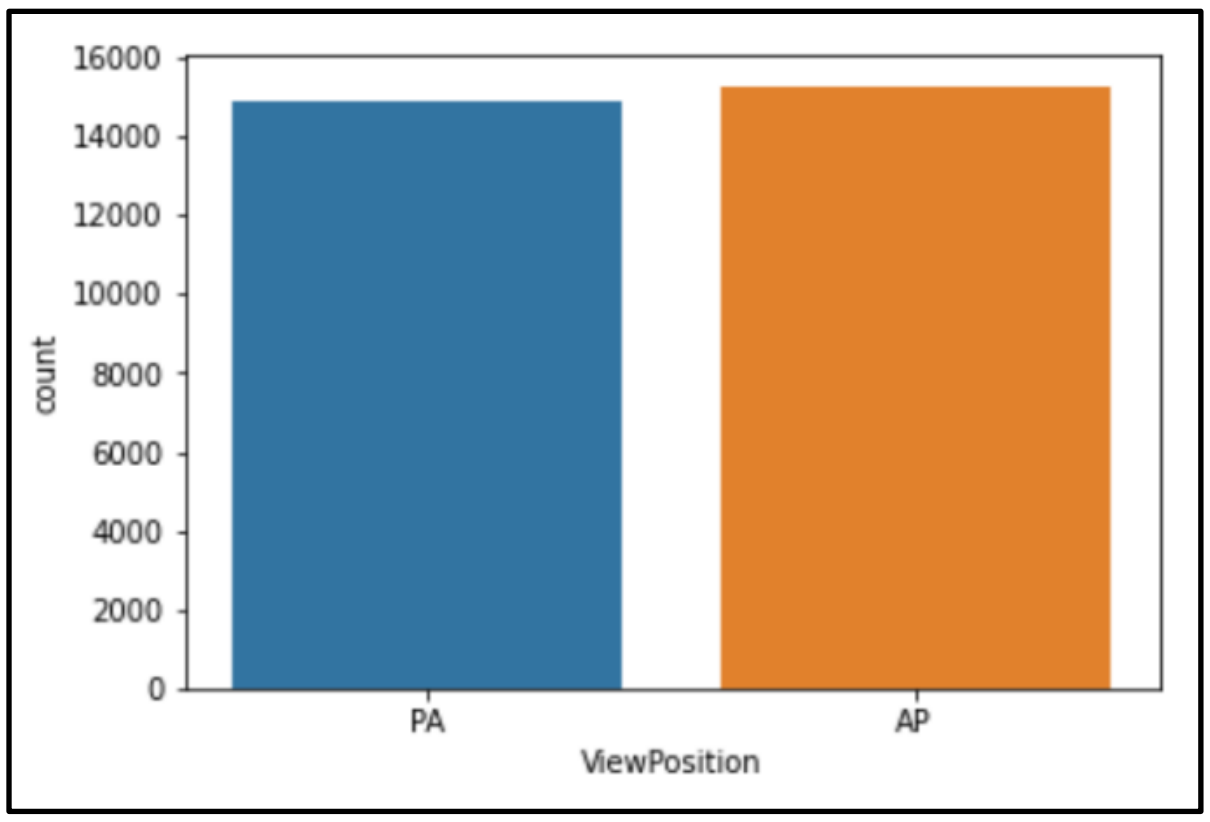

Fig. 11. Count plot of PA & AP

Inference: 'AP' and 'PA' View Positions doesn't seem to have much of a difference in terms of the amount of data.

*2.2.6.1.4. Bounding box info*

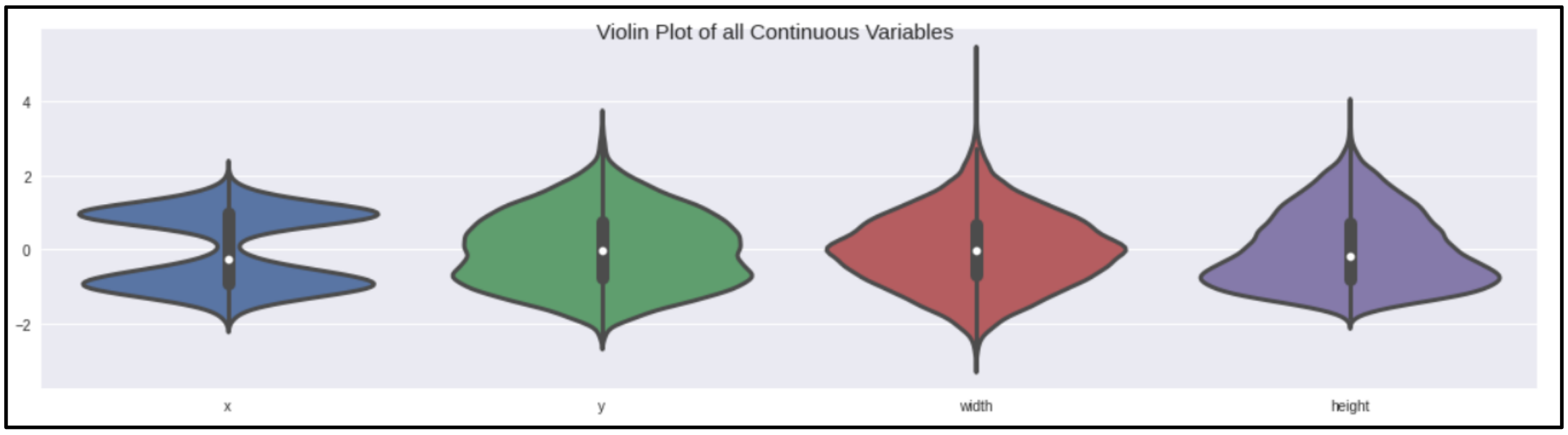

Fig. 12. Violin plot of a bounding box

*2.2.6.2. Bivariate Analysis*

*2.2.6.2.1. View Position concerning Target*

1. 'PA' view position is highly imbalanced for the 'Target' attribute while the 'AP' view position is balanced. Fact patients should be imaged in an upright PA position, as AP views are less useful and only reserved for very ill patients who are unable to stand erect.
2. The inference is that patients that are imaged in an AP position are those that are more ill, and therefore more likely to have contracted pneumonia. As the absolute split between AP and PA images is about 50-50, the above consideration is extremely significant.

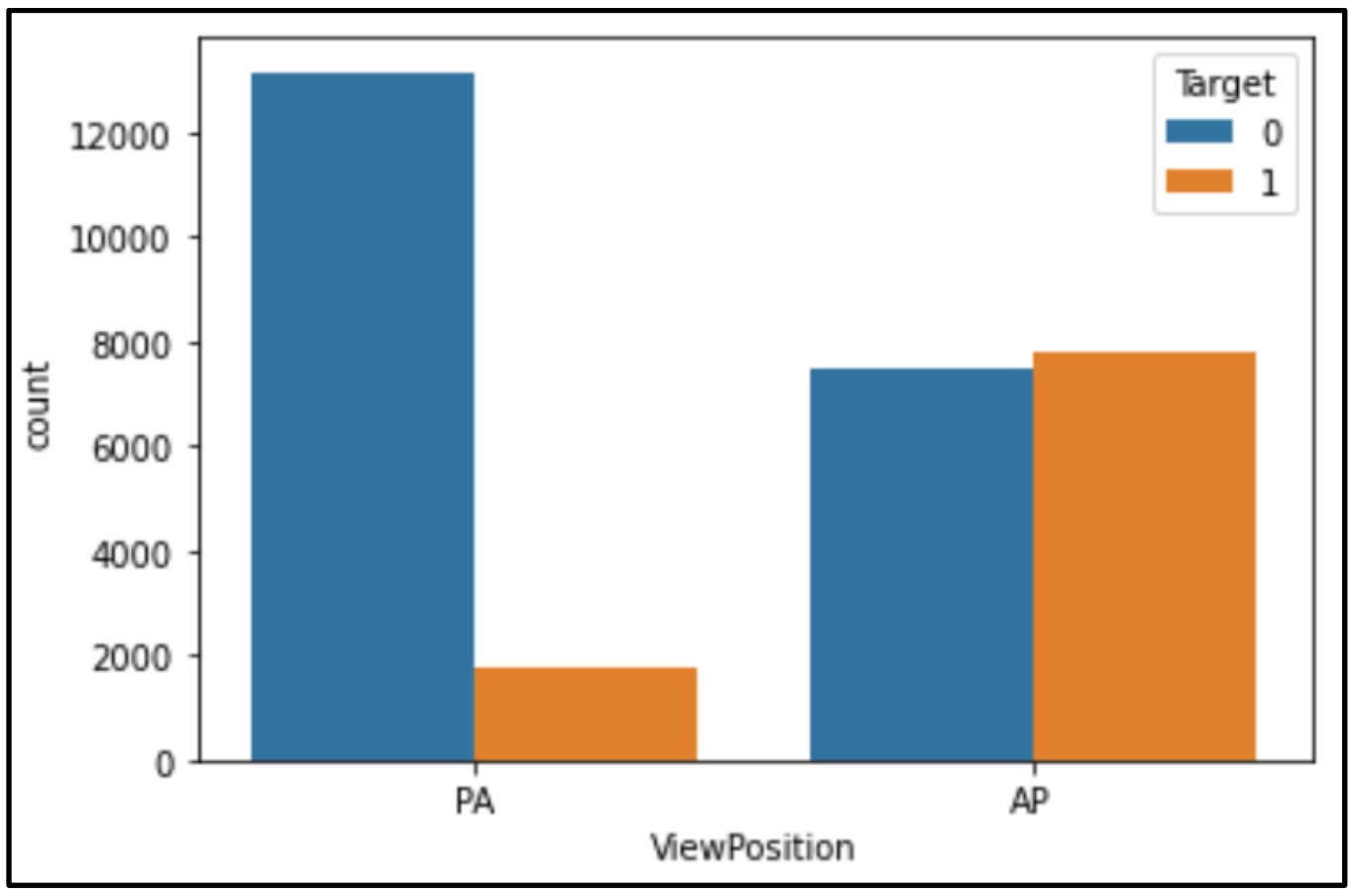

Fig. 13. Stack plot of PA & AP

Inference: 'PA' view position is highly imbalanced concerning the 'Target' attribute while the 'AP' view position is balanced.

*2.2.6.2.1. Patient's Sex to Target*

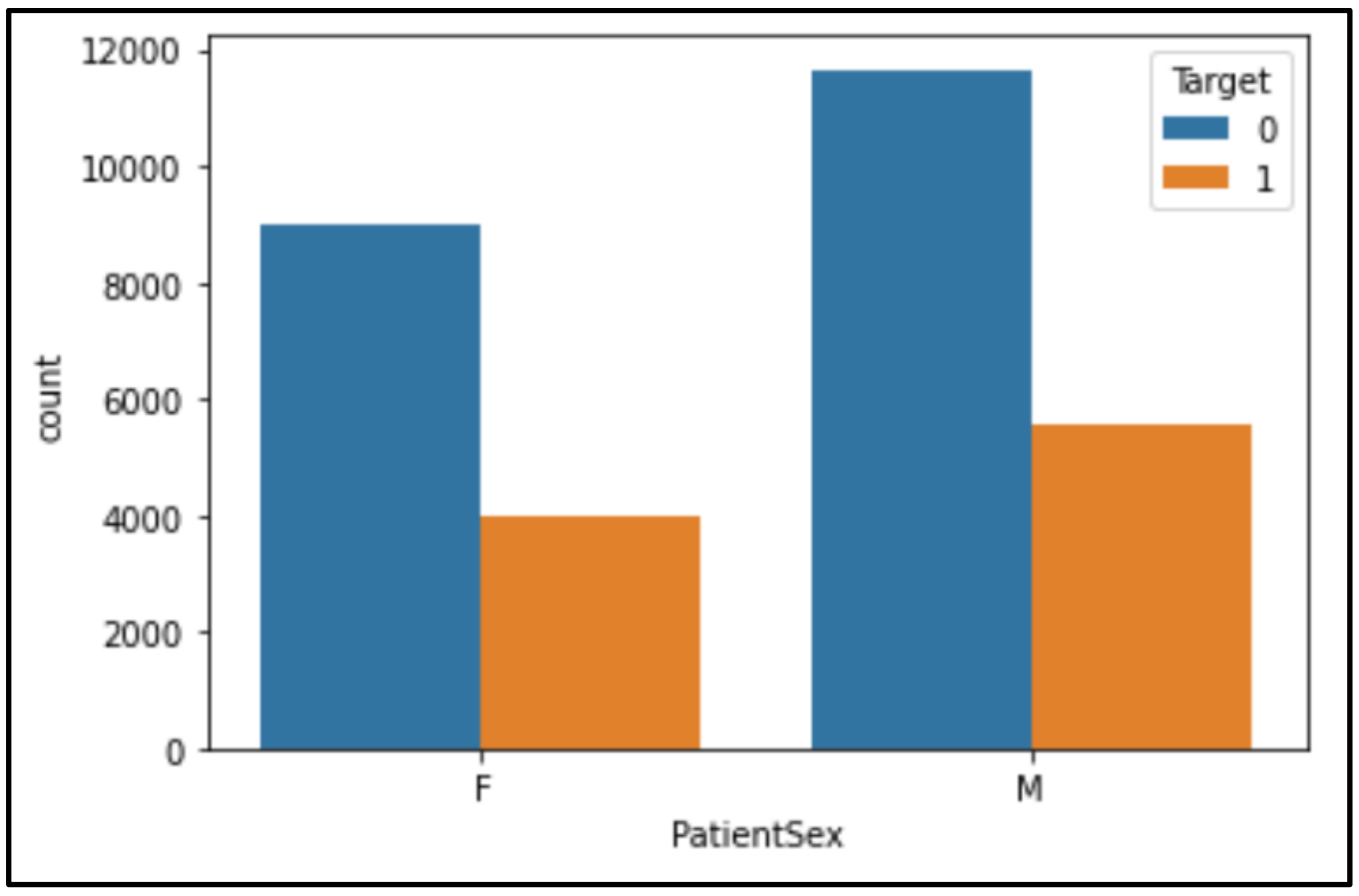

Fig. 14. Stack plot of male & female

Inference: The ratio between the Patient's Sex concerning Target seems to be the same between males and females when compared to the number of data present. But what is noticeable is that more male patients are positive with pneumonia as compared to female patients.

### *2.2.7. Augmentation*

1. The dataset has a large number of images. Among the three classes, they are imbalanced. With the help of augmentation, it can improve the training. The different types of augmentation are applied here:
   a. Rotation_range,
   b. Width_shift_range,
   c. Height_shift_range,
   d. Shear_range, Zoom_range & Normalization

### *2.2.8. Image Filtering*

1. One of the image histogram equalization processing called Clahe has been applied to the X-ray images as an experiment.

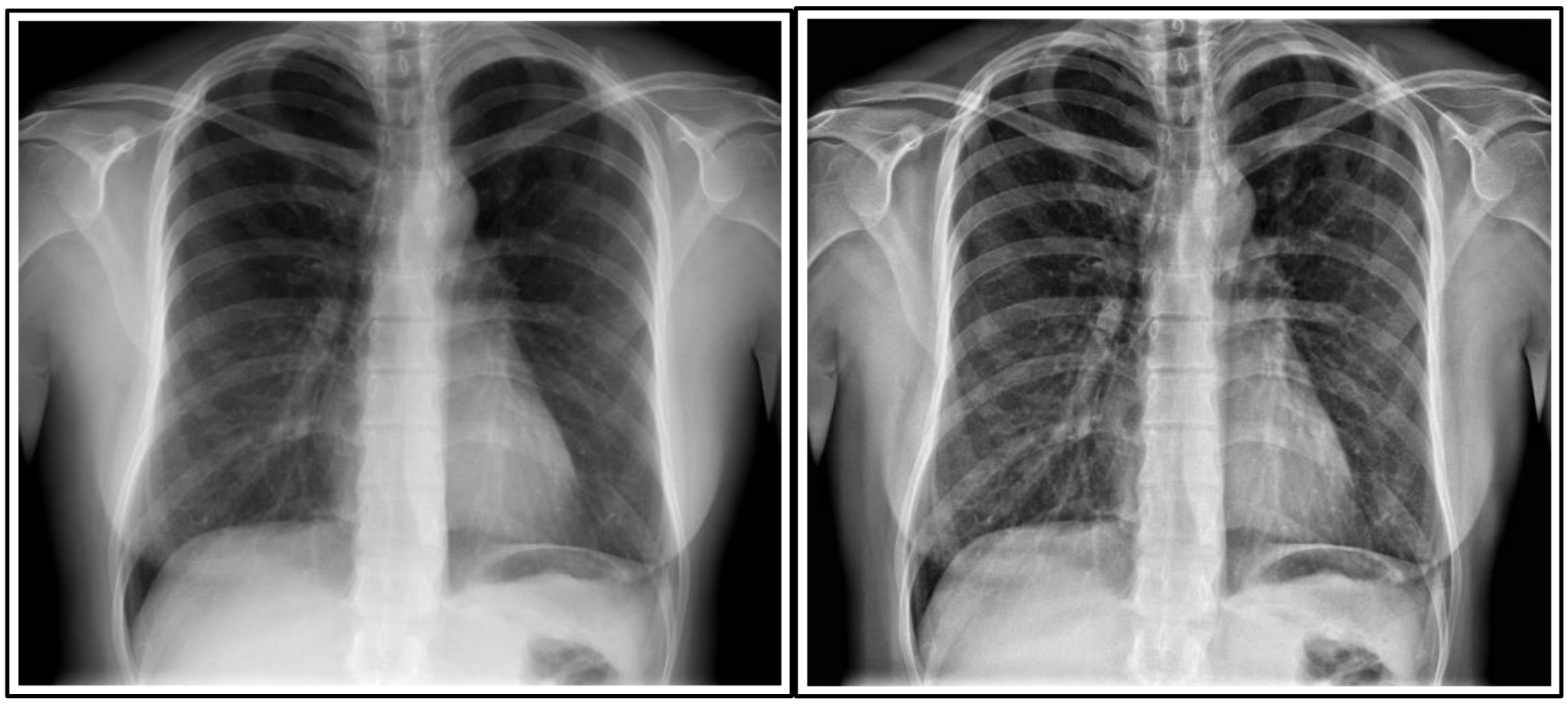

Fig. 15. (a). Without Clahe filter; (b). With Clahe filter

2. In image enhancements, during filtering, there are a lot of noises detected in the image. Therefore, applying Clahe can try to prevent those noises and try to improve the quality of the image.

## 3. Methodology

### *3.1. Overview*

The study started with the literature survey and analysis of all the available data. The inferences gathered from data and survey was used to create a basic CNN model. From the basic CNN model, the study moved to the transfer learning models. The transfer learning model layers were not trained instead the pre-trained weights from imagnet were only used. Three different transfer learning models were trained on 3017 stratified samples with input images downsized to 224 X 244. Later an ensemble of three transfer learning models was tried with max voting. The results were good but only half of our problem was getting solved. Hence, we moved to advanced computer vision algorithms like masked RCNN. This yielded the results we were trying to achieve.

### *3.2. CNN*

For the base model, a CNN architecture was used to perform the binary classification of X-rays. From the data, the duplicates were first removed. After the removal of duplicates, 26684 unique x-rays remained. Further, the class labels were manually encoded. All the entries having no lung opacity or normal were considered as 0 and those having lung opacity were labelled as 1. The original class column was then removed. Then stratified sampling on the data was performed using Cochran's formula. This resulted in a total of 2203 x-rays with the label distribution being 1707 for 0 class and 496 for 1 class. After stratification, the 2203 x-rays were extracted from the main dataset. These images were extracted using the pydicom library and were converted to pixel arrays for further processing. The chest x-ray matrix thus obtained had to shape as (2203, 1024, 1024). For the same 2203 patients, the corresponding labels were extracted. The shape of the label matrix was changed to (2203, 1) for ease of working with CNN architecture. The pixel information was checked for the chest x-ray matrix and max and min values were found to be 255 and 0 respectively. The images were then resized from 1024 to 224 pixels. For this, the resize function was used from skimage library. After resizing the max and min values of pixels became 1.000 and 0. The data was then split into train and test with a 30% split. The X_train and X_test were reshaped to (1542, 224, 224, 1) and (661, 224, 224, 1) for making them compatible with the CNN algorithm. The base model architecture had 5 Convolutional layers along with MaxPooling layers. The first two convolutional layers had 16 and 32 filters with size (3,3). Whereas the remaining three convolutional layers had 64 filters with size (3,3). The activation function used for each convolutional layer was Relu. For the MaxPooling layers, a pool size of (2,2) was used. After the final MaxPooling layer, the output was flattened for providing it as input to fully connected layers. A dense layer with 512 neurons and a Relu activation function was used. The final output layer had 1 neuron for binary classification with activation function Sigmoid.

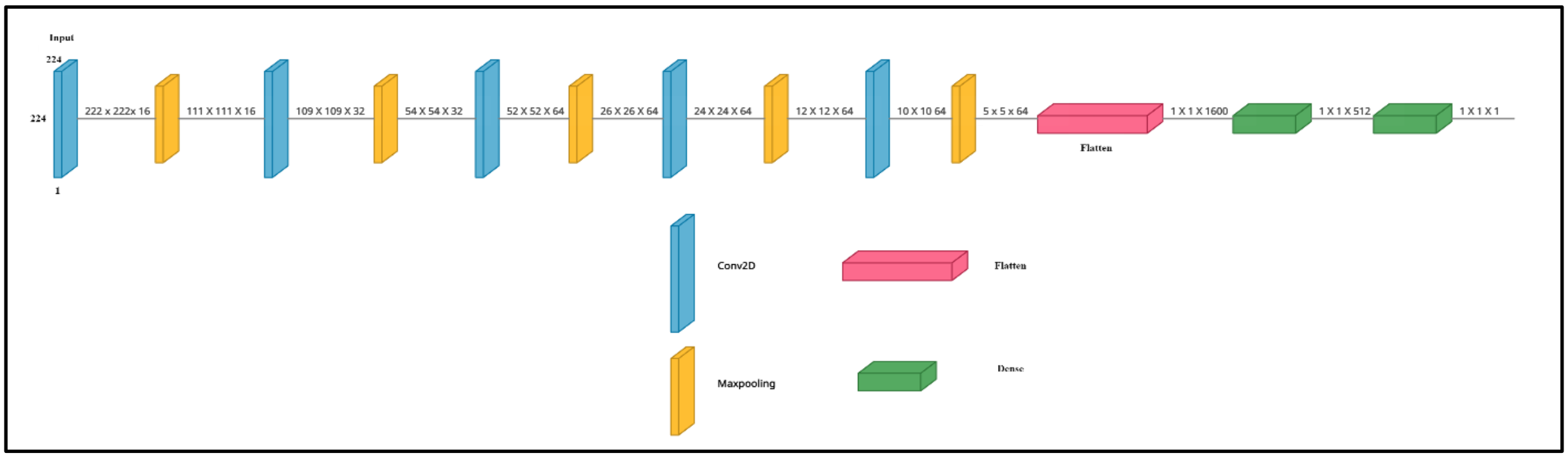

Fig. 16. Architecture of CNN

### 3.3. *Transfer Learning*

While performing the EDA, the RSNA dataset has highly unbalanced labels. Therefore, the downsample technique has been applied. In this experiment, the RSNA dataset's third label - No opacity/normal was removed. The callbacks are applied to save the best model using val_accuracy and reduce learning rate metrics.

#### 3.3.1. *Model I*

1. The first models' purpose is to detect Pneumonia from the chest X-ray.
2. In this experiment, two types of separate CNN's transfer learning models - ResNet50 and InceptionV4 (Modified) have been implemented.
3. The models' shape - (224x224), used only Imagenet weights in ResNet50 and included custom top layer- GlobalAveragePooling and dropout of 20%.
4. The parameters - 20 epochs (ResNet50), 30 epochs (InceptionV4) and 64 batch size, sigmoid activation, and Adam optimizer with default learning rate.

#### 3.3.2. *Model II*

1. The second model is for experimental purposes, and its objective is to find the view position (PA-AP) of the chest.
2. This model is trained on the RSNA dataset, and its test data is used for validation.
3. In the EDA, similar to model I, the ap-pa parts are also unbalanced, so downsample technique was applied to balance the features.
4. This model is performed on the transfer learning model- EfficientNetB0.
5. The parameters and hyper-parameters are applied the same as the model I.

#### 3.3.3. *Model III*

1. The third model's purpose is to detect pneumonia with the clahe filter.
2. This model is trained on modified backbone architecture- InceptionNetV4 without pre-trained weights with scaling of -1 to 1, the shape of (224x224), top layer - GlobalAveragePooling and dropout- 0.25.
3. The parameters and hyper-parameters are almost the same as model I.

### 3.4. *Mask R-CNN*

Mask RCNN is an evolution of Faster RCNN. It has an additional Segmentation Mask for every detected object. Along with the object detection and localization, there is also one FCN that takes ROI as input and predicts a segmentation mask.

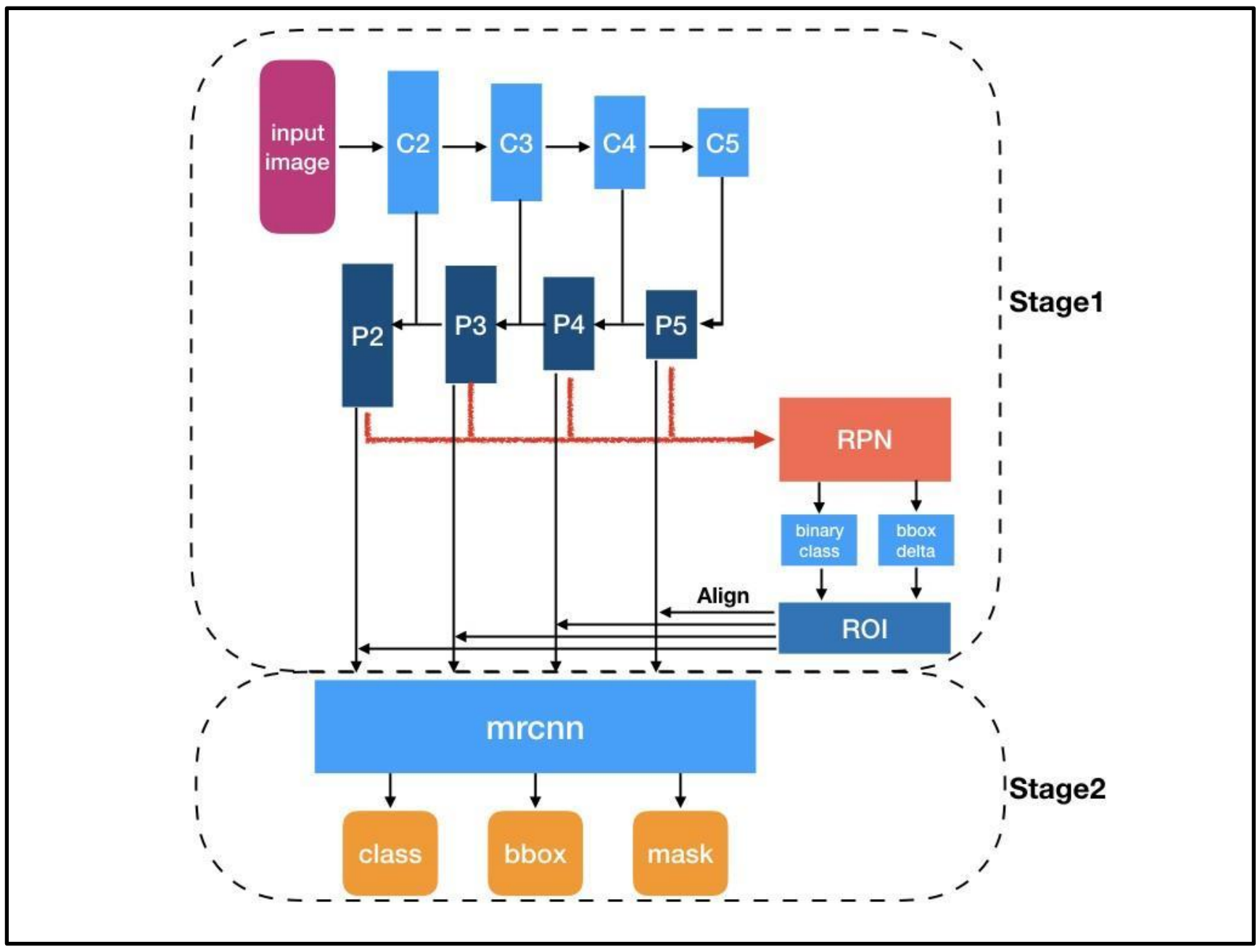

Fig. 17. The model architecture of R-CNN

1. For implementing Mask RCNN on the RSNA dataset, Matterport's implementation of Mask R-CNN was used. Two backbones were tried – Resnet50 and Resnet101. The image dimensions were reduced to (256,256) and image augmentation was applied. The model training was performed in a jupyter environment with TensorFlow-GPU enabled.
2. Initially, the model was trained and validated on a stratified sample of 3017 images. Model 1 was fitted for 16 epochs with a learning rate of 0.006. The first 2 epochs were trained with a higher learning rate to make the learning process quicker. Finally, all the layers were trained for 200 steps per epoch.
3. For Model 2, the number of images was increased to 4836 based on stratified sampling. Image augmentation was not used in this model. The model was trained with a resnet50 backbone. A decrease in loss was noticed after increasing the images.
4. For Model 3, the backbone was changed to resnet101. With this, the classification loss was further reduced.
5. Finally, for Model 4, the number of image samples were increased to 9068 images. However, when the number of images is increased, the bounding box detection performance is reduced and loss increases to 0.5216.

### *3.5. YOLOv3*

Yolo stands for You only look once. It is a real-time object detection system. It is known to be faster than other similar CNN architectures. However, it retains the accuracy of the detection. The Yolo architecture divides an image into multiple regions. The bounding boxes and their probabilities are predicted for each region of the image. For the RSNA dataset, we trained a Yolo v3 implementation. The Yolo v3 architecture uses Darknet-53 as the backbone feature extractor. It is more efficient compared to other backbones such as Resnet101.

The model was trained only on the pneumonia class images. The train set was 5410 images and the validation set was 602 images. Image augmentation was applied to the dataset. A learning rate of 0.001 was used. The model was compiled for 900 iterations. It took 6 hours to compile the model with a 6 GB GPU.

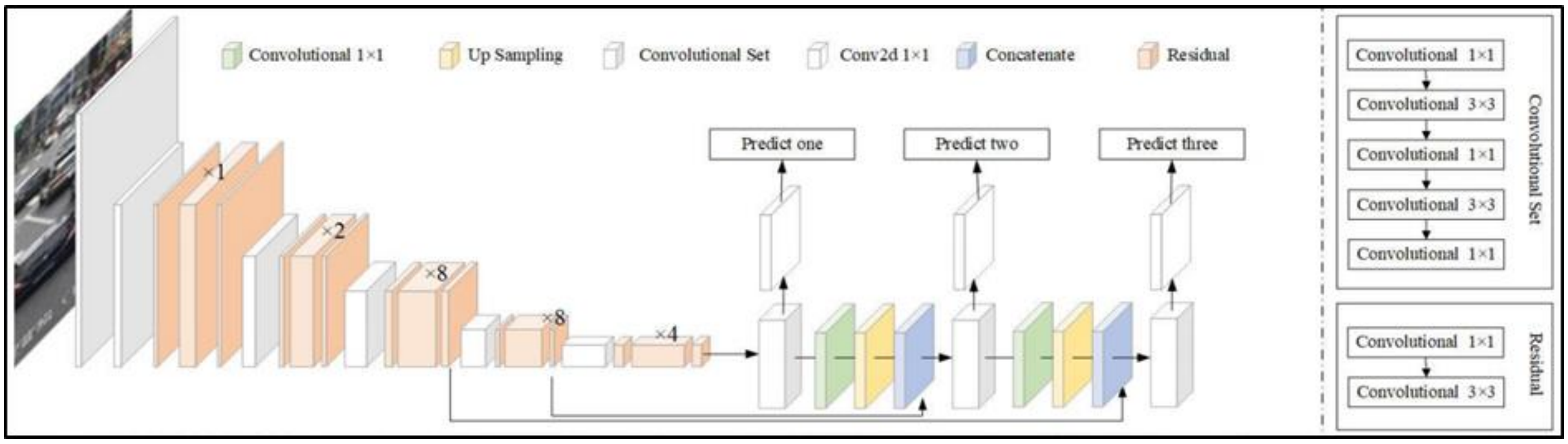

Fig. 18. Yolov3 Structure

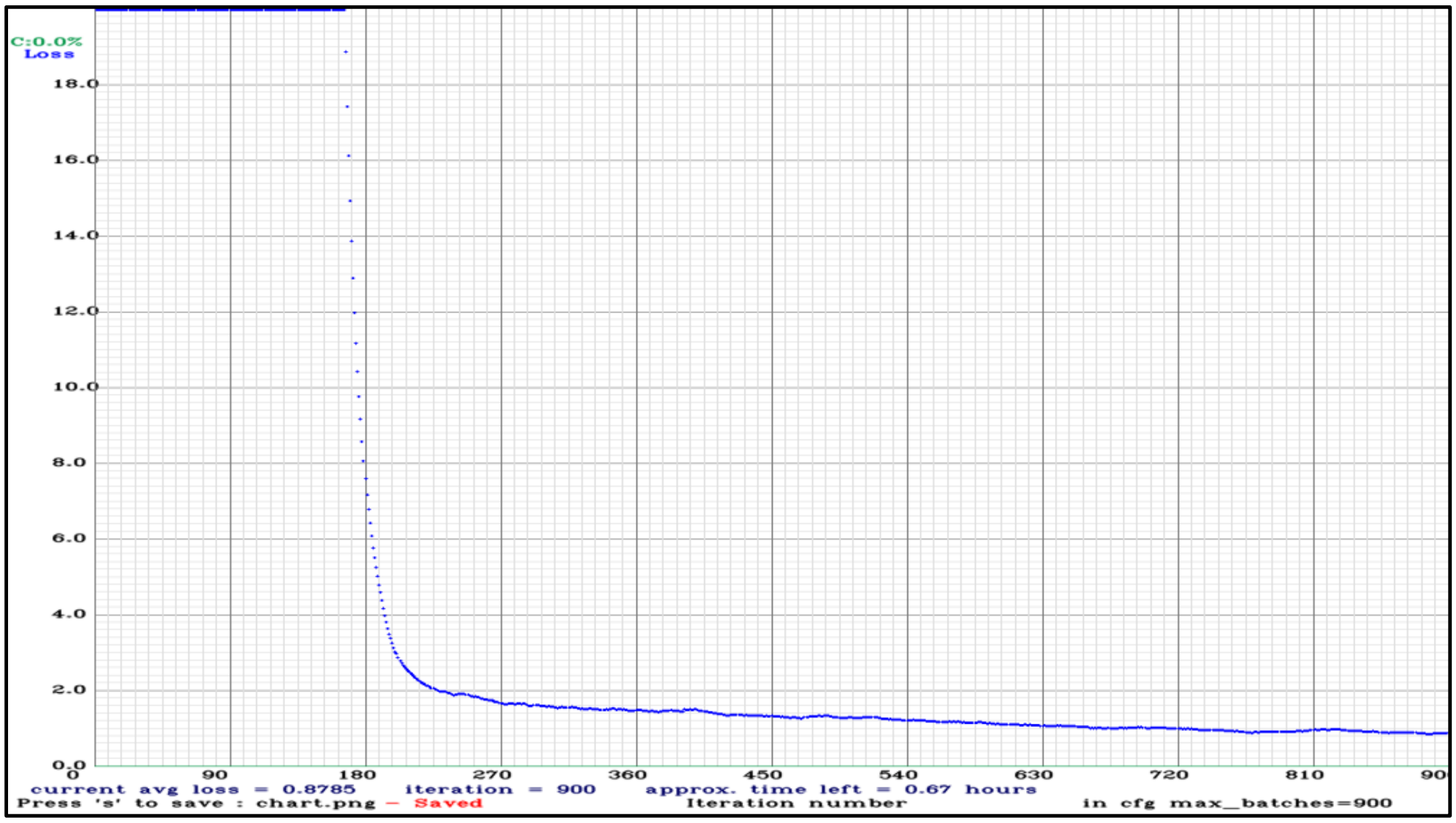

Fig. 19. mAP score chart

The above chart shows the average loss value concerning the iterations. The average loss of class and bounding box prediction was 0.8785. The mean average precision (mAP) of the model was reported as 0.32.

## 4. Results

### *4.1. Early Results*

The model was compiled using binary cross-entropy as the loss function. RMSProp was used as the optimizer with a learning rate of 0.001. The metric used for evaluation was accuracy. For fitting the model 25 epochs were used along with a batch size of 16. The validation data was also provided for the fit. The model was run on a GPU based

Jupyter environment for faster processing. The final result at the 25th epoch was - train accuracy 95%, validation accuracy 79%.

### *4.2. Interim Results*

The above result shows the basic model is overfitting. Hence to avoid or improve overfitting, three transfer learning models with a learning rate of 0.0001 experimented. Except for the learning rate no other hyper-parameters were changed. The transfer learning models tried were Xception, Inception V3 and EfficientNetB7. All the models were fitted for 30 epochs of batch size 128 with the early stopping of 5 patience.

The results are as follows: -

Table 1. Transfer learning results

| **Sr.No.** | **Model Name** | **Training Accuracy** | **Validation Accuracy** |
|---|---|---|---|
| 1. | Xception | 77.14% | 75.87% |
| 2. | InceptionV3 | 75.97% | 73.77% |
| 3. | EfficientB7 | 75.07% | 75.76% |

### *4.2. Transfer Learning Results*

#### *4.2.1. Model I*

ResNet50 got a training score of 92%, validation score of 90% while InceptionV4 got a training score of 96% and a validation score of 94%.

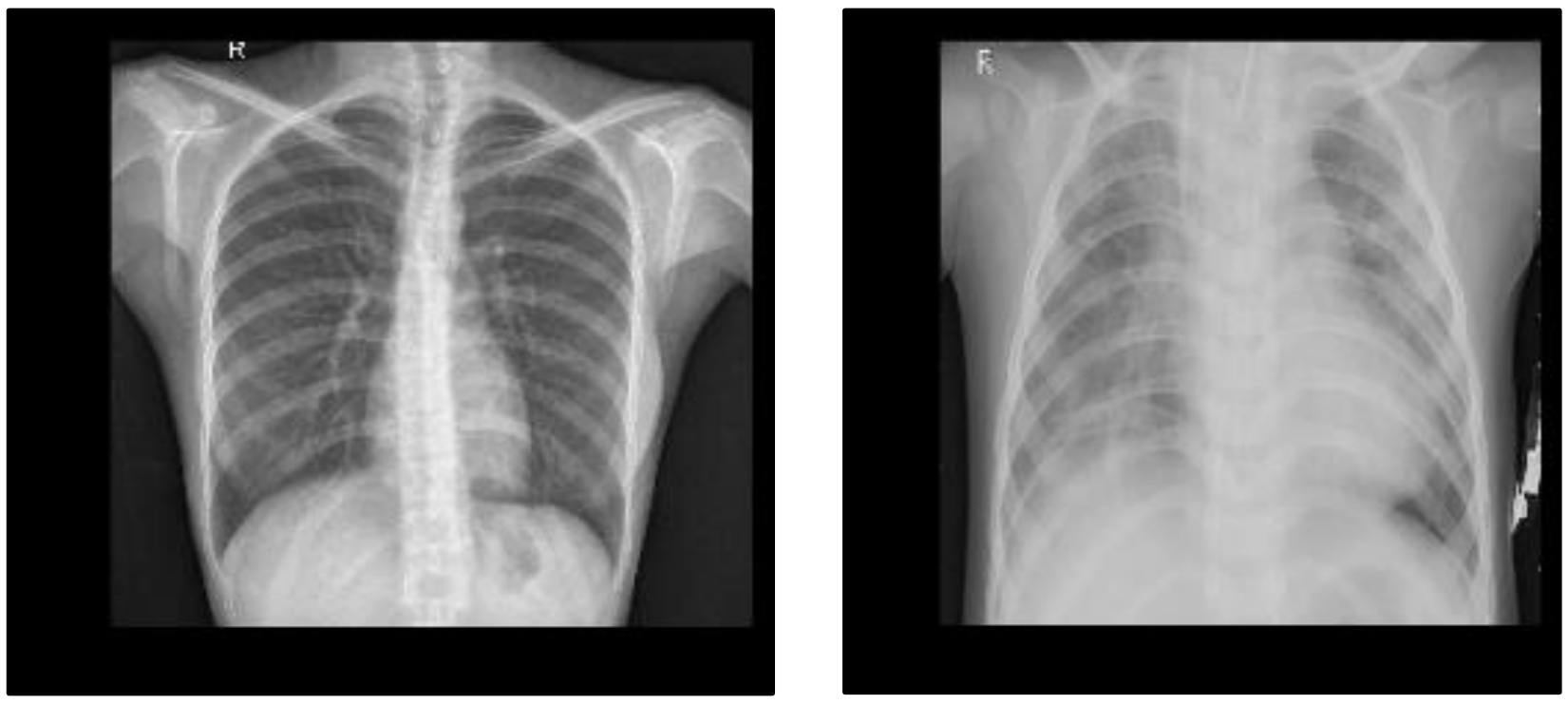

Fig. 20. (a). Actual: Lung Opacity, Predicted : Normal (ResNet50); (b). Actual: Normal, Predicted : Normal (InceptionNetv4)

#### *4.2.2. Model II*

This model is performed well in this experiment. The validation score of this model is 98.4%, with a minimum loss of 5%. The classification report is AP - 98% and PA - 99%. This model is not effective when it comes to classifying Pneumonia detection.

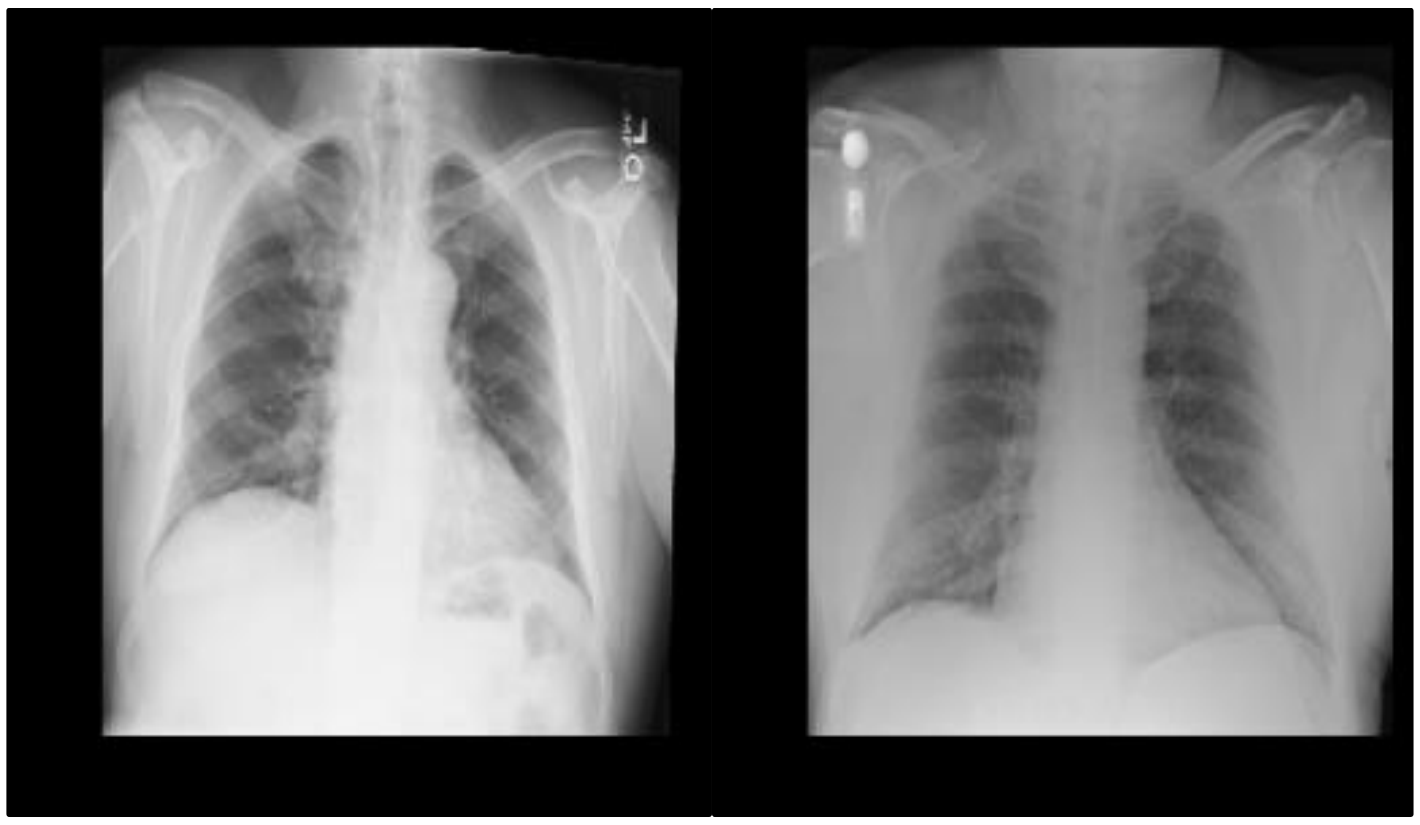

Fig. 21. (a). Actual: PA, Predicted : PA; (b). Actual: AP, Predicted: AP

### *4.2.3. Model III*

The clahe model improved a little bit in the positive class while the accuracy of the normal class declined.

### *4.3. Mask R-CNN Results*

The losses for classification and localization are being calculated separately. The classification loss is based on binary cross-entropy while localization loss is based on a mean squared error of the bounding box. Model three here gives the best results based on losses but the MAP score of model 1 is 0.15.

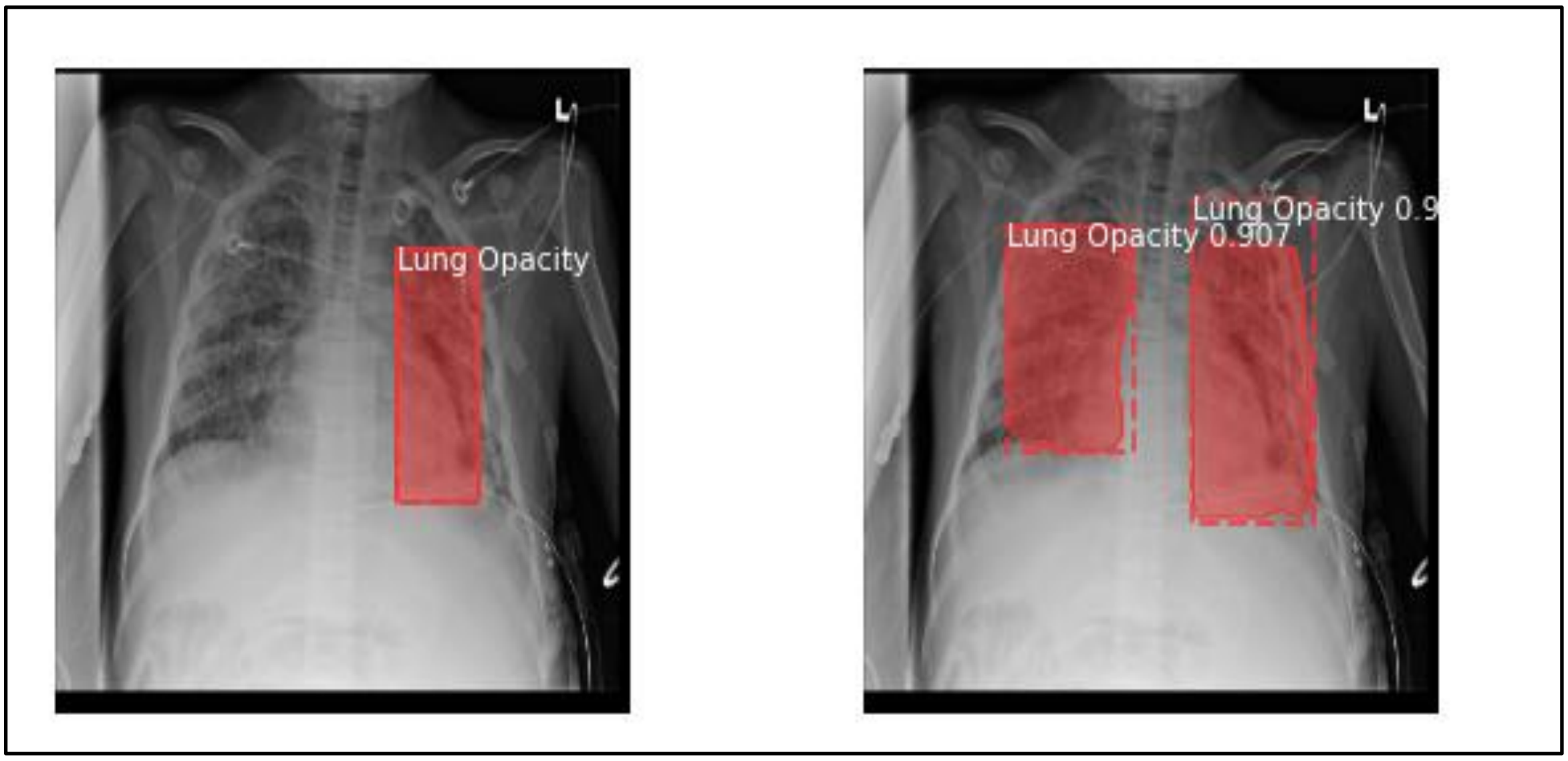

Fig. 22. (a). Actual; (b). Predicted

Table 2. Models' result

| **No. of Images** | **Model** | **Backbone** | **Results** | |
|---|---|---|---|---|
| | | | **Classification Loss** | **Localization Loss** |
| 3017 | Model 1 | resnet50 | 0.2635 | 0.3982 |
| 4836 | Model 2 | resnet50 | 0.2071 | 0.2784 |
| 4836 | Model 3 | resnet101 | 0.1734 | 0.2818 |
| 9068 | Model 4 | resnet101 | 0.214 | 0.5216 |

### *4.4. Yolov3 Results*

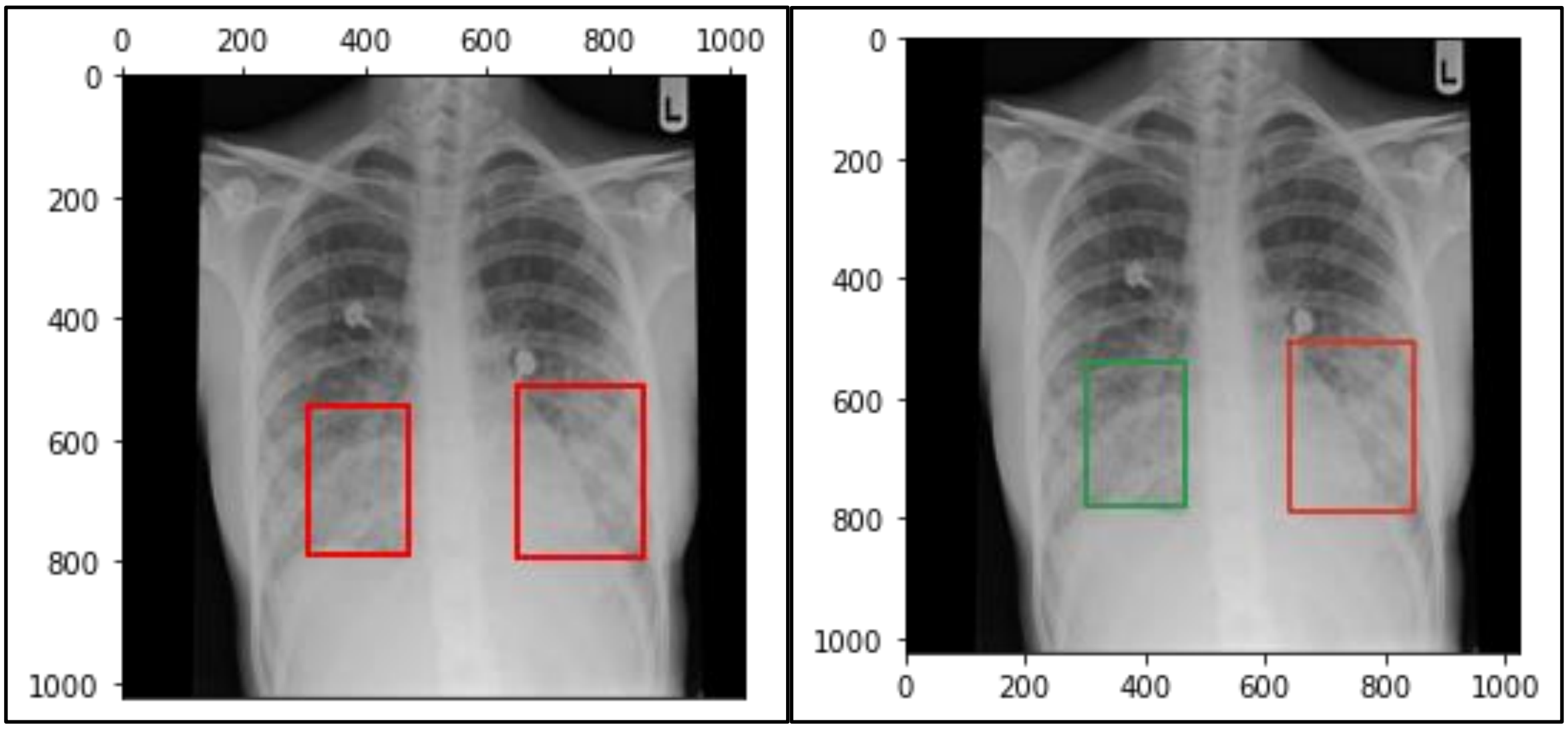

Fig. 23. (a). Actual; (b). Predicted

The model showed good results with a bounding box threshold of 0.1 as depicted above for one of the patients. The final mAP of the model was 0.32 which is better than the RSNA challenge top score of 0.25.

## 5. Conclusion & Discussion

We tried various models to achieve our goal of building a model with minimal computational resources so that faster and accurate results can be brought out in the healthcare industry, especially the imaging side as it is very computationally heavy. This led us to the YoloV3 model which without any hyperparameter tuning gave us results close to the benchmark. This opens the possibility of running the model for further iterations as the loss still keeps decreasing.

### *5.1. Closing Reflections*

1. Less data and a constant learning rate limit the model from achieving higher accuracies.
2. Stratified sampling of data for training, gives the model less opportunity of learning the needed feature in turn leaving the model with fewer accuracies as here only 1/3rd data is of the positive class.
3. Using transfer learning CNN methods to get better accuracy.
4. Implementation of the Yolov3.

### *5.2. Limitation*

A few key limitations of our approach are as below:

1. The computational power is limited.

2. The classes are unbalanced, and one of the classes - Not Normal/Pneumonia is identified as unknown in the dataset.
3. The AP-PA model is not effective when it comes to classifying Pneumonia detection.
4. Clahe filter didn't improve on normal x - rays, it only works accurately on the pneumonia positive x - rays.
5. To avoid running out of available memory, the batch and subdivisions for Yolo architecture were selected such that only one image is loaded in the GPU for each iteration.
6. At 900 iterations the loss was still decreasing. So, if a greater number of iterations can be processed with larger batch size, the training time will be considerably lesser.
7. Without a GPU it is not feasible to run the Yolov3 architecture.
8. The model may improve further after image pre-processing.
9. Different variants of image augmentation can be tried.
10. The size of predicted bounding boxes can be scaled down for better accuracy.

### *5.3. General Discussion*

This whole process of data analysis and model building gave us a big learning experience of how to install libraries to debug the errors that came along with implementations. The major learning was how to implement on limited resources and to choose the right model for the same. Through various experiments, Hyperparameters and their effects on models were understood. As the model building takes its own sweet time to fit the data and learn, using the right parameters/hyper-parameters and guiding the model for better learning is something we would like to do differently.

Further, the basic CNN models were found good for chest x-ray classification. Also, the advanced object detection architectures require lots of pre-work before training. This includes creating a suitable environment with all the necessary dependencies. The newer architectures such as Yolov3 can give higher mAP compared to transfer learning-based ensemble models which are resource-intensive.